\documentclass{article}
\usepackage{spconf,amsmath,graphicx,amsfonts,amssymb,subfigure,multirow,cite,booktabs}

\title{FRAME-WISE STREAMING END-TO-END SPEAKER DIARIZATION WITH NON-AUTOREGRESSIVE SELF-ATTENTION-BASED ATTRACTORS}
\name{Di Liang$^{1,2}$, Nian Shao$^{1,2}$, Xiaofei Li$^{2,\ast}$ \thanks{*Corresponding Author}}
\address{$^{1}$Zhejiang University, Hangzhou, China \\ $^{2}$Westlake University \& Westlake Institute for Advanced Study, Hangzhou, China}
\begin{document}
\ninept
\maketitle
\begin{abstract}
This work proposes a frame-wise online/streaming end-to-end neural diarization (FS-EEND) method in a frame-in-frame-out fashion. To frame-wisely detect a flexible number of speakers and extract/update their corresponding attractors, we propose to leverage a causal speaker embedding encoder and an online 
non-autoregressive self-attention-based attractor decoder. A look-ahead mechanism is adopted to allow leveraging some future frames for effectively detecting new speakers in real time and adaptively updating speaker attractors. The proposed method processes the audio stream frame by frame, and has a low inference latency caused by the look-ahead frames. Experiments show that, compared with the recently proposed block-wise online methods, our method FS-EEND achieves state-of-the-art diarization results, with a low inference latency and computational cost. 
\end{abstract}
\begin{keywords}
streaming speaker diarization, end-to-end, self-attention, online inference, frame-in-frame-out
\end{keywords}
\section{Introduction}
\label{sec:intro}

Speaker diarization is to identify speakers as well as their active intervals within a given audio recording. It answers the question ``who spoke when'' in a multi-speaker environment \cite{anguera2012speaker}. Speaker diarization has a wide range of audio and video applications, such as intelligent conferencing \cite{kang2020multimodal} and medical systems \cite{finley2018automated}. Joint speaker diarization and automatic speech recognition (ASR) has been proven to improve the ASR performance \cite{shafey2019joint}. 
The multi-stage framework used to be the mainstream technique \cite{dimitriadis2017developing, zhang2019fully}, where a diarization system consists of four successive stages: speech activity detection (SAD), speech segmentation, speaker embedding extraction, and clustering.
Although clustering-based methods could handle a flexible number of speakers, they cannot deal with speaker overlap because each frame is bound with one speaker.
Also, it cannot directly utilize the diarization tags presented in training data.

End-to-end approaches are proposed to overcome the above problems \cite{fujita2019end1, fujita2019end2, horiguchi2020end, han2021bw}.
Fujita et al. \cite{fujita2019end1} propose an end-to-end neural diarization (EEND) model with bidirectional LSTM (Bi-LSTM) \cite{zhou2016attention} for speaker embedding extraction.
Permutation invariant training (PIT) \cite{yu2017permutation} is conducted for resolving the label ambiguity problem.
Transformer encoder is used to replace Bi-LSTM in \cite{fujita2019end2}. Since the fixed network output dimension of EEND model limits the number of speakers, Horiguchi et al. \cite{horiguchi2020end} propose a LSTM-based encoder-decoder attractor (EDA) mechanism to obtain a flexible number of attractors, and diarization is performed by taking the inner product between the attractors and the frame-wise speaker embeddings. Transformer decoder is used instead of LSTM to generate attractors for better diarization performance in \cite{rybicka2023attentionbased, chen2023attentionbased}.

In some online applications such as multi-party human-robot-interaction or real-time subtitling of video conference, the diarization system is required to process audio streaming and recognize/react to the speaking person in real time. To this aim, some online speaker diarization methods \cite{han2021bw, wang2018speaker, xue2021online1} have been proposed. Based on the offline EEND-EDA method \cite{horiguchi2020end}, a block-wise EEND-EDA (BW-EEND-EDA) method is proposed in \cite{han2021bw}. Speaker embeddings and attractors are calculated incrementally per 10 s audio block, which leads to a 10 s inference latency. 
Xue et al. \cite{xue2021online1,xue2021online2} design an inference-stage speaker-tracing buffer (STB) for online diarization. STB stores the previous input frames and diarization results. 
For a new chunk input, the new frames are stacked with frames in the buffer to perform a new diarization. The new diarization results are permuted according to the previous diarization results stored in the buffer to keep speaker order the same.
STB can achieve a low inference latency by setting a small chunk size, such as 1 second. However, this requires additional computations since all frames in the buffer need to be re-fed to the diarization network for a single chunk inference. In \cite{horiguchi2022online}, unsupervised clustering is introduced to attractor-based EEND to enable diarization of unlimited numbers of speakers. Moreover, a variable chunk-size training (VCT) mechanism is adopted to mitigate the diarization error at the very beginning part of recordings.

Different from the block-wise online methods \cite{han2021bw, xue2021online1, xue2021online2,horiguchi2022online}, this work proposes a frame-wise online/streaming EEND method, named FS-EEND. The proposed model mainly consists of a causal speaker embedding encoder and an online attractor decoder. The online decoder extracts attractors frame-wisely, and is realized with a non-autoregressive self-attention network. Two dimensions of self-attention are designed: the causal time self-attention attends to the attractor of previous frames for the same speaker to maintain the attractor consistency along time, while the cross-attractor self-attention attends to the attrator of other speakers to better distinguish speakers.  
The major difficulties for frame-wise streaming diarization lie in detecting new speakers in real time and adaptively updating speaker attractors. To mitigate these difficulties, a look-ahead mechanism implemented by 1-dimensional convolution is adopted to allow leveraging some future frames for making more confident decisions. In addition, an embedding similarity loss is designed to guide emmbedding convergency. The proposed frame-wise streaming method processes the audio stream frame by frame, and has a low inference latency caused by the look-ahead frames. Experiments show that, compared with the block-wise online methods \cite{han2021bw, xue2021online2}, the proposed frame-wise streaming method achieves better diarization performance, with a lower inference latency and computational cost. 

\section{METHOD}
\label{sec:method}

Speaker diarization is defined as a multi-class detection task where the timestamps of the active speakers should be recognized for a given audio recording. We denote an audio recording in the time-frequency domain as a sequence of LogMel feature vectors: $\boldsymbol{X} = (\boldsymbol{x}_1, \ldots, \boldsymbol{x}_1, \ldots, \boldsymbol{x}_{T})$ with $ \boldsymbol{x}_t \in \mathbb{R} ^ {F}$, where $T, F$ stands for the number of frames and the dimension of feature vector, respectively. The corresponding speech activity label for $S$ speakers is represented as $\boldsymbol{Y} = (\boldsymbol{y}_1, \ldots,\boldsymbol{y}_t,\ldots, \boldsymbol{y}_{T})$ with $\boldsymbol{y}_t \in \{0, 1\}^{S}$.

\subsection{Speaker appearance order}
\label{sec:order}
In offline scenarios, the speaker order is arbitrary, and the diarization model can be trained with the PIT loss \cite{horiguchi2020end}. By contrast, in streaming online scenarios, the speakers appear one by one, and the speaker order should be set according to the speaker appearance order. 

\begin{figure}[t]
  \centering
  \includegraphics[width=0.7\linewidth]{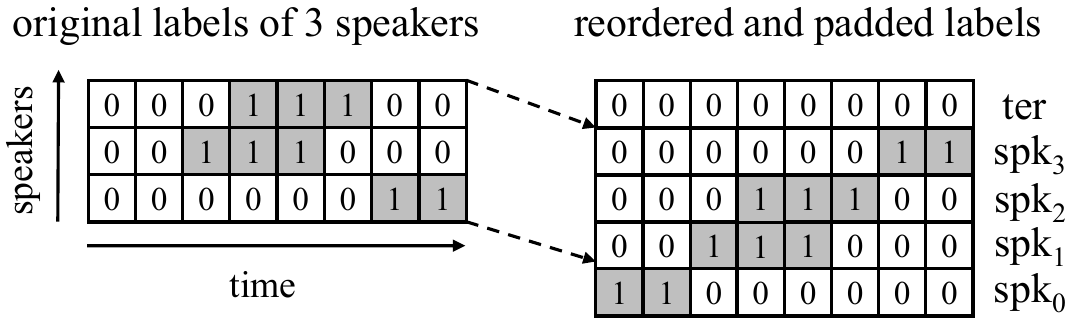}
  \caption{An example of speaker labelling according to the speaker appearance order. $\text{spk}_{\text{0}}$ represent an extra non-speech speaker. 'ter' stands for the termination of active speakers. } 
  \label{fig:label_transf}
  \vspace{-1em}
\end{figure}

Fig.~\ref{fig:label_transf} shows the speaker order for the proposed method. One extra speaker is added for the non-speech frames, acting as a speech activity detector. At the training stage, the non-speech speaker is set as spk$_\text{0}$, and the actual $s$ speakers are successively set from spk$_\text{1}$ to spk$_s$ following their appearance order. To determine the number of speakers, we concatenate a speaker termination marker (zero label) on the top of active speaker labels. 

\begin{figure}[t]
  \centering
  \includegraphics[width=0.80\linewidth]{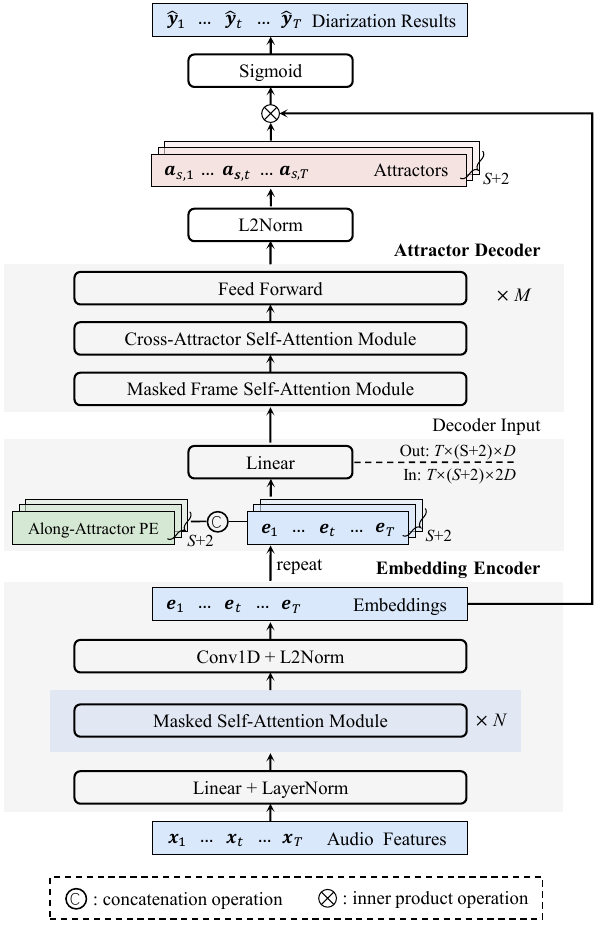}
  \caption{Architecture of the proposed FS-EEND system.} 
  \label{fig:model}
  \vspace{-1em}
\end{figure}

\subsection{Architecture of frame-wise streaming EEND}
\label{sec: architecture}
The proposed frame-wise streaming diarization system is shown in Fig.~\ref{fig:model}, which is mainly composed of a speaker embedding encoder, a look-ahead convolutional layer and an attractor decoder. The speaker embeddings are frame-wisely extracted. The attractors are generated on the fly and updated frame-wisely. In this work, both the speaker embeddings and attractors are normalized to have a unit L2-norm. L2-norm normalization is widely used for representation learning \cite{li2022atst} to improve the stability of embeddings.

\subsubsection{Causal speaker embedding encoder}
 The speaker embedding sequence $\boldsymbol{E}=(\boldsymbol{e}_1, \ldots,\boldsymbol{e}_t, \ldots, \boldsymbol{e}_T)$, $\boldsymbol{e}_t \in \mathbb{R} ^ {D} $ is first causally extracted from the input feature sequence $\boldsymbol{X} = (\boldsymbol{x}_1, \ldots, \boldsymbol{x}_1, \ldots, \boldsymbol{x}_{T})$ with a frame-wise linear layer and a masked transformer encoder \cite{vaswani2017attention}
\begin{align}
  \boldsymbol{E} &= \text{MaskedTransformerEncoder}(\boldsymbol{X}).
\end{align}
We apply a lower triangular binary matrix to each self-attention module in the transformer encoder. Such binary masks ensure that each frame-wise embedding is calculated only from its previous contexts, avoiding the use of future information.

\subsubsection{Look-ahead with 1-dimensional convolution}
 For most of applications, a certain diarization latency is tolerable. Exploiting some future information is helpful for improving the quality of speaker embeddings and attractors. Especially, the birth of a new speaker is subtle to be detected when the new speaker appears in only few frames. Therefore, we adopt a look-ahead mechanism to exploit some future frames and make more confident decisions, by applying 1-dimensional convolution along the time dimension. The latency (number of future frames) is determined by the convolution kernel size. 
Then the embeddings are L2-norm normalized.

\subsubsection{Online attractor decoder}

The speaker embeddings are fed into an online attractor decoder to extract frame-wise attractors. The decoder is designed to detect a new speaker/attractor immediately when the new speaker appears. Then, the attractors of existing speakers will be updated as more and more speaker embeddings collected along the increase of time. Different from the existing EEND methods \cite{han2021bw, xue2021online1, xue2021online2,horiguchi2022online} that use a LSTM-based attractor extractor, we design a self-attention-based attractor decoder in this work, as the self-attention mechanism has a longer memory to not lose the information of a previously appeared speaker. The task of attractor extraction for speaker diarization is different from other sequence generation tasks, such as speech recognition and machine translation. The multiple attractors do not have a certain ordered semantic dependency as the text sequence, and thus it is better to process the multiple attractors in parallel. Therefore, in this work we propose a non-autoregressive attractor decoder to simultanously output multiple attractors. A maximum number (e.g. $S=4$ in this work) is set for the number of attractors/speakers. This will limit the use of the proposed method when the number of speakers is larger than the preset maximum number. However, it is shown in \cite{horiguchi2022online} that, even if there is no a preset maximum number in the model, the model (whether online or offline EEND) still cannot handle the case when the number of inference speakers is larger than the number of training speakers. And the problem of unlimited number of speakers needs to be solved in other ways.    

The architecture of the proposed attractor decoder is shown in Fig.~\ref{fig:model}. The decoder data has two sequence dimensions, one is the time-frame dimension, and the other is the attractor dimension. The decoder outputs an attractor sequence for each frame.

\noindent\textbf{Decoder input}.
At one frame, the speaker embedding of the current frame is input to the decoder to update the corresponding attractor. All the $S+2$ attractors use the same repeated speaker embedding, and the decoder decides whether the speaker embedding is used for attractor updating.  
The input of multiple attractors need to be distinguished from each other. To this end, we concatenate the speaker embedding with a positional encoding (PE) \cite{vaswani2017attention} along the attractor dimension, followed by a linear layer as the final encoder input. 

\noindent\textbf{Decoder architecture}.
For deriving the attractor $\boldsymbol{a}_{s,t}$ of the $s$-th speaker at frame $t$, two information sources need to be referred to: i) the attractor of previous frames for the same speaker, i.e.  $\boldsymbol{a}_{s,t^{\prime}}, t'< t$. Seeing the attractor of previous frames can maintain the attractor consistency of one speaker along time. To refer to previous frames, a masked frame self-attention (MFSA) module is designed; ii) the attractor of other speakers at frame $t$, i.e.  $\boldsymbol{a}_{s^{\prime},t}, s'\neq s$. By seeing other attractors, the distance between attractors may be enlarged to better distinguish different speakers. For this purpose, a cross-attractor self-attention (CASA) module is designed.  
Our self-attention block is designed following the standard transformer encoder  \cite{vaswani2017attention}. The MFSA and CASA modules are two multi-head self-attention modules  along the time-frame and attractor dimensions, respectively. Note that MFSA also uses the binary masks to ensure the causality, as is done in the causal speaker embedding encoder. After the self-attention modules, a feed-forward module is used. The three modules are realized with a residual connection and layer normalization as in \cite{vaswani2017attention}. $M$ self-attention blocks are cascaded. 

\noindent\textbf{Decoder output}.
The decoder output vectors are also normalized to have a unit L2-norm as the final attractors $\boldsymbol{a}_{s,t}$. The speech activity of all speakers are computed by taking the inner product between the frame-wise attractors $\boldsymbol{A}_{t}=[\boldsymbol{a}_{1,t},\ldots,\boldsymbol{a}_{s,t},\ldots,\boldsymbol{a}_{S,t}]$ and the speaker embedding $\boldsymbol{e}_t$ as $\hat{\boldsymbol{y}}_{t} = \sigma ( \boldsymbol{A}_t^{\intercal} \boldsymbol{e}_t )$.

\subsection{Model training}


A binary cross entropy (BCE) loss is used for network training for the estimated speech activities $\hat{\boldsymbol{Y}}$ and the ground truth $\boldsymbol{Y}$ (as shown in the right table of Fig.~\ref{fig:label_transf}), $\mathcal{L}_d=\frac{1}{T} \sum^{T}_{t=1} \text{BCE} ( \hat{\boldsymbol{y}}_t , \boldsymbol{y}_t)$.

\textbf{Embedding similarity loss}.
The clustering effect of the frame-wise speaker embeddings have salient impacts on diarization performance. An embedding similarity loss is introduced to restrict the distribution of the embeddings. 
We constrain the cosine similarity between each pair of embeddings ($\langle \boldsymbol{e}_j, \boldsymbol{e}_k \rangle, j, k \in \{1, \ldots, T\}$) to be consistent with the cosine similarity between their corresponding diarization labels ($\langle \boldsymbol{y}_j, \boldsymbol{y}_k \rangle, j, k \in \{1, \ldots, T\}$). The mean square error between the two cosine similarities is minimized 
\begin{align}
  \mathcal{L}_e = \frac{1}{T\times T} \sum_{j=1}^{T} \sum_{k=1}^{T} \text{MSE}(\langle \boldsymbol{e}_j, \boldsymbol{e}_k \rangle, \langle \boldsymbol{y}_j, \boldsymbol{y}_k \rangle). \label{eq: l_e}
\end{align}
%
This loss is suitable for various situations, including two frames from the same, different, and overlapped speakers.

The total loss for training is the sum of the diarization loss and the embedding similarity loss, i.e. $\mathcal{L} = \mathcal{L}_d+\mathcal{L}_e$.

\section{Experiments}
\label{sec:experiments}

\subsection{Dataset}

The simulated speech mixtures are generated from Switchboard Cellular (Part 1 and 2) and 2005-2008 NIST Speaker Recognition Evaluation (SRE) datasets according to the recipe presented in \cite{fujita2019end2}. There are 4054 speakers in total, where 3244, 405 and 405 of them are used for training, validation and test, respectively. Noise from MUSAN corpus \cite{snyder2015musan} and Simulated Room Impulse Response \cite{ko2017study} are applied to the synthesized dataset. We evaluate the proposed method on the real telephone conversations from the CALLHOME corpus \cite{martin2001nist} as well. 250 realistic recordings are used to finetune our model, leaving 242 recordings for test.

\subsection{Experimental settings}

We released our codes on our website\footnote{https://github.com/Audio-WestlakeU/FS-EEND}. Following the input generation recipe presented in \cite{horiguchi2020end}, the input features are 345-dimensional LogMels. We use four-stacked masked transformer encoders in embedding encoder (i.e., $N$=4) and two-stacked attractor decoders (i.e., $M$=2).  For both the encoders and decoders, the number of hidden units and heads are set to 256 and 4, respectively. The kernel size and padding in the look-ahead 1-dimensional convolution are set to 19 and 9, respectively. In this way, each frame aggregates information from  both the previous and subsequent 9 frames, resulting in a latency of 1 s.
We first train the model on a 2-speaker dataset including 100,000 simulated mixtures for 100 epochs and then finetune the 2-speaker model using the simulated mixtures with a varying number of speakers (1, 2, 3 or 4 speakers) for 50 epochs. Finally, we finetune the model with the CALLHOME adaption data for 100 epochs and evaluate on CALLHOME test data (148, 74, and 20 recordings for 2, 3, and 4 speakers, respectively). 

The system is evaluated through diarization error rate (DER) with a collar tolerance of 250 ms. As presented in Sec.~\ref{sec:order}, during training, the speaker identities are determined by the speaker appearance order. At inference, it may fails to enroll (generate an attractor) for one new speaker when the speaker appears at the first time, which leads to a wrong speaker order. However, this wrong speaker order does not affect the diarization performance, as the speaker can still be correctly detected when the speaker is enrolled afterward. Hence, for performance evaluation, we do not follow the speaker appearance order, instead, the optimal mapping between predictions and ground truth are used, as the same evaluation metric for offline EEND \cite{horiguchi2020end}. 

\subsection{Experimental results}

\subsubsection{Ablation study}
\begin{table}[t]
  \caption{DERs (\%) of ablation studies.}
  \footnotesize
  \label{tab:DERs1}
  \centering
    \renewcommand\arraystretch{1.0}
\tabcolsep0.05in
  \begin{tabular}{lcccc}
    \toprule
    \centering
    \multirow{2}*{Methods} & \multicolumn{4}{c}{Number of speakers} \\ 
    ~ & 1 & 2 & 3 & 4  \\
    \midrule
\textbf{FS-EEND} (prop.) & \multicolumn{1}{c}{{0.6}} & \multicolumn{1}{c}{{5.1}} & \multicolumn{1}{c}{{11.1}} & \multicolumn{1}{c}{15.8}\\
\quad  with appearance order & \multicolumn{1}{c}{{0.6}} & \multicolumn{1}{c}{{6.8}} & \multicolumn{1}{c}{15.1}  & \multicolumn{1}{c}{{21.6}} \\
\quad  w/o L2-normalization  & \multicolumn{1}{c}{0.5} & \multicolumn{1}{c}{5.3} & \multicolumn{1}{c}{11.8} & \multicolumn{1}{c}{17.0} \\ 
 \quad  w/o embedding similarity loss  & \multicolumn{1}{c}{1.0} & \multicolumn{1}{c}{6.4} & \multicolumn{1}{c}{13.2} & \multicolumn{1}{c}{19.6}\\
 \quad  w/o look-ahead  & \multicolumn{1}{c}{0.8} & \multicolumn{1}{c}{6.7} & \multicolumn{1}{c}{13.6} & \multicolumn{1}{c}{19.9}\\
\bottomrule
\end{tabular}
\end{table}
\begin{figure}[t]
    \centering
    \includegraphics[width=0.35\linewidth]{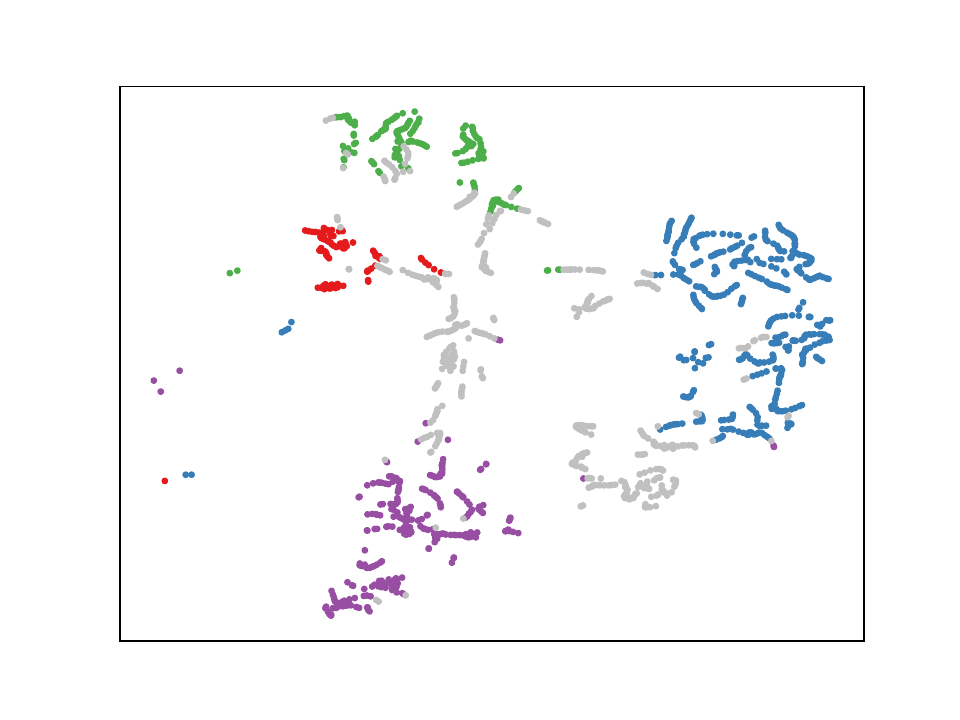}
    \hspace{0.02\linewidth}
    \includegraphics[width=0.35\linewidth]{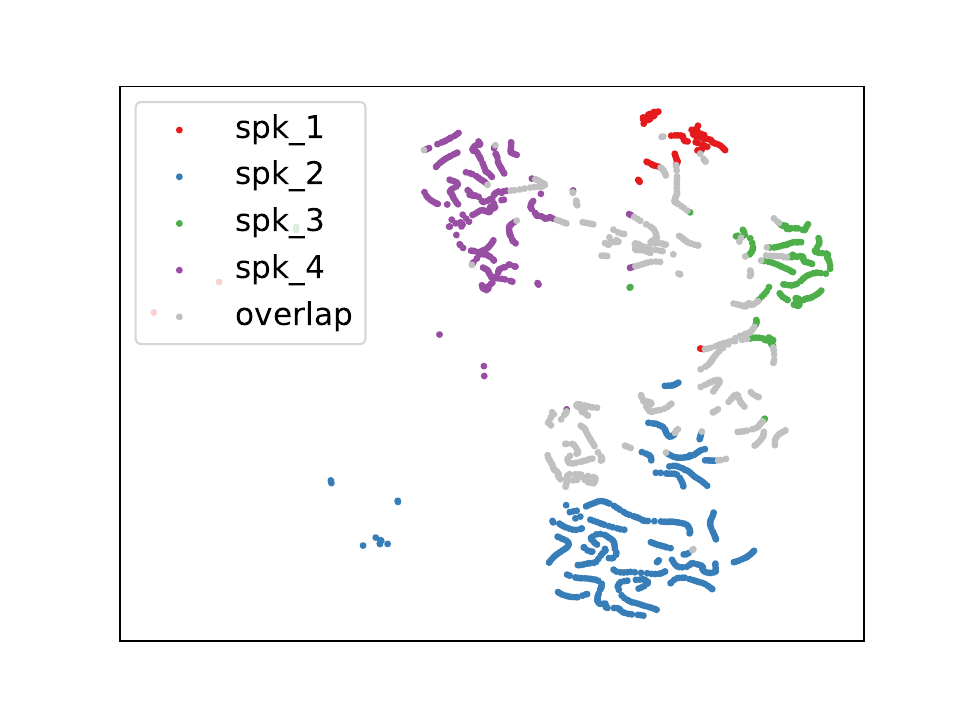} 
    \\
    {(a)}\hspace{0.35\linewidth}{(b)}
    \\ \vspace{0.2em}
    \includegraphics[width=0.35\linewidth]{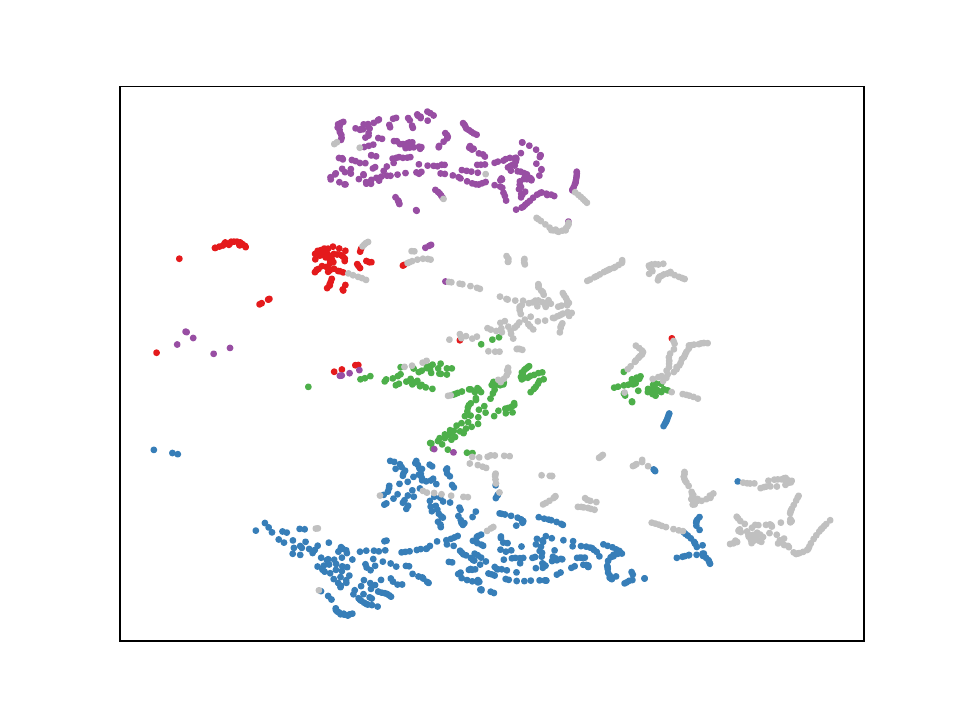}
    \hspace{0.02\linewidth}
    \includegraphics[width=0.35\linewidth]{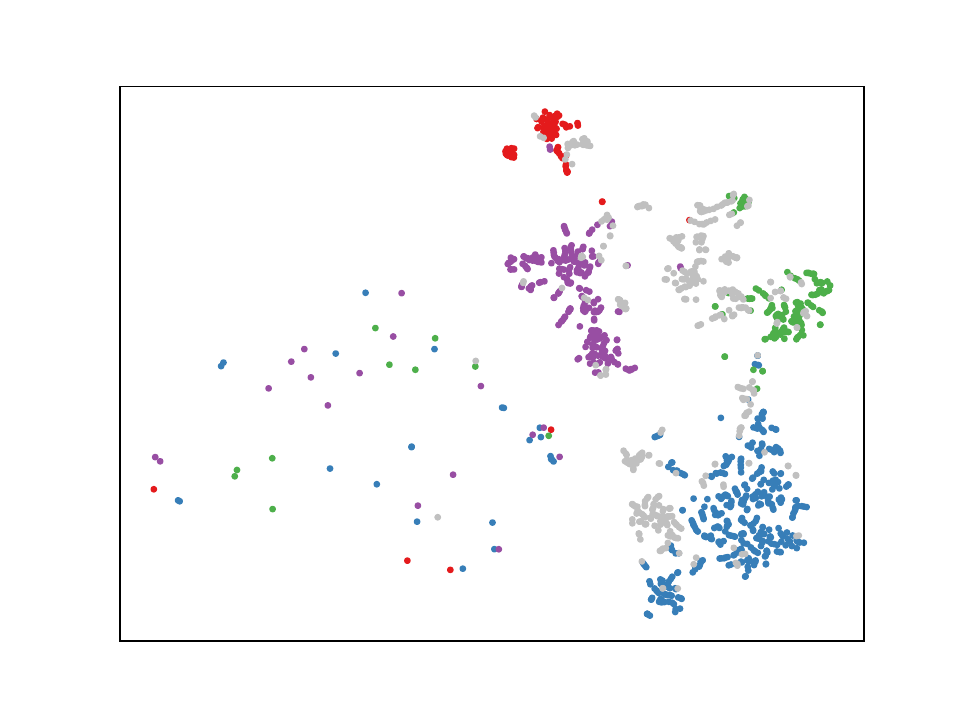}   
    \\
    {(c)}\hspace{0.35\linewidth}{(d)}
    \centering
    \caption{t-SNE visualization \cite{van2008visualizing} of embeddings in 2-dimensional space. (a) the proposed FS-EEND, (b) without L2-normalization, (c) without embedding similarity loss and (d) without look-ahead. }
    \vspace{-1.5em}
    \label{fig:visual}
\end{figure}

Table~\ref{tab:DERs1} shows the results of some ablation studies on the simulated data. When the performance is evaluated \emph{with appearance order}, the DERs get higher. The performance degradation indicates the enrollment failure rate when the speakers appear at the first time. The enrollment failure rate increases along the increasing of speaker numbers. 
The effectiveness of L2-normalization, embedding similarity loss, and look-ahead are validated by the results shown in Table~\ref{tab:DERs1}, as the DERs get higher when one of them is not used. To not use look-ahead, we change the look-ahead convolution to be causal with a kernel size of 10. 

Fig.~\ref{fig:visual} visualizes and compares the speaker embeddings obtained with and without L2-normalization, embedding similarity loss and look-ahead. Comparing Fig.~\ref{fig:visual} (b) and (a), it can be seen that L2-normalization has no significant effect in terms of the aggregation of the same speaker and the discrimination of different speakers. Comparing Fig.~\ref{fig:visual} (c) and (a), it can be seen that the embedding similarity loss leads to a better clustering of embeddings for the same speakers, and also makes the embeddings of overlapping speech distributed in between the single-speaker embeddings. Comparing Fig.~\ref{fig:visual} (d) and (a), it can be seen that look-ahead helps to remove some outliers.

\subsubsection{Results on simulated data}

\begin{table}[t]
    \caption{DERs (\%) and RTF on the simulated data. All the methods are implemented by ourselves 
    under the same dataset.}
    \footnotesize
    \label{tab:DERs2}
    \centering
    \renewcommand\arraystretch{1.0}
    \tabcolsep0.04in    
\begin{tabular}{lcccccc}
    \toprule
    \centering
    \multirow{2}*{Methods} & \multirow{2}*{RTF} & \multirow{1}{*}{latency } &  \multicolumn{4}{c}{Number of speakers}\\ 
    ~ & ~ & (s) & 1 & 2 & 3 & 4  \\
    \midrule
    Offline EEND-EDA \cite{horiguchi2020end}   & 0.006 & -  & 0.4 & 4.0 & 9.9  & 14.1 \\
    EEND-EDA+FLEX-STB \cite{xue2021online2}    & 0.028 & 10  & 0.7 & 4.7 & 13.0 & 17.1 \\
    EEND-EDA+FLEX-STB \cite{xue2021online2}    & 0.223  & 1 & 1.9 & 6.7 & 15.1 & 19.6 \\
    \textbf{FS-EEND} (prop.)                   & 0.026 & 1  & 0.6 & 5.1 & 11.1 & 15.8 \\ 
    \bottomrule
\end{tabular}
\vspace{-1em}
\end{table}

We compare the proposed FS-EEND method with the offline EEND method \cite{horiguchi2020end} and the recently proposed online EEND-EDA with FLEX-STB method \cite{xue2021online2} on simulated data with a flexible number of speakers. The buffer size of FLEX-STB is 100 s, and the chunk size (system latency) is set to 10 s or 1 s. The results are shown in Table \ref{tab:DERs2}. It is not surprising that the online methods perform worse than the offline method. FLEX-STB with 10 s chunk size achieves lower DERs than FLEX-STB with 1 s chunk size, at the cost of a much larger inference latency. With 1 s inference latency, the proposed method achieves substantially lower DERs than FLEX-STB with 1 s chunk size, and noticeably lower DERs even than FLEX-STB with 10 s chunk size for most of cases. The good performance of the proposed FS-EEND attributes to the proposed self-attention attractor decoder, which has longer meomery than LSTM for aggragating embedding information and extracting attractors. It attributes to the proposed strategies of L2-normalization, embedding similarity loss and look-ahead as well. 

To compare the computational complexity of different methods, we evaluate the real-time factor (RTF) on an AMD EPYC 7742 64-Core CPU Processor using 1 thread. The RTFs shown in Table \ref{tab:DERs2} are calculated on simulated 4-spk data. The offline method computes the speaker embedding once per frame, and computes the attractors once for the entire recording using the LSTM encoder-decoder. The proposed FS-EEND computes the attractors frame-wisely. Therefore, the RTF of the proposed method is larger than the one of the offline method. The block-wise online method, i.e. FLEX-STB, needs to compute the speaker embedding $\frac{\text{buffer size}}{\text{chuck size}}$ times per frame due to the overlap of blocks. As a result, the RTF of FLEX-STB with 1 s chunk size is one order of magnitude larger than the one of FS-EEND.

\subsubsection{Results on CALLHOME data}

\begin{table}[t]
    \caption{DERs (\%) on CALLHOME data.}
    \footnotesize
    \label{tab:DERs3}
    \centering
\begin{tabular}{lccccc}
    \toprule
    \multirow{2}{*}{Methods} & \multirow{1}{*}{latency} & \multicolumn{3}{c}{Number of speakers} \\
    ~ & (s) & 2 & 3 & 4 \\
    \midrule
    Offline EEND-EDA \cite{horiguchi2020end}  & -  & 7.7  & 13.7 & 22.4 \\
    BW-EEND-EDA \cite{han2021bw}              & 10 & 11.8 & 18.3 & 25.9 \\                                                  
    EEND-EDA+FLEX-STB \cite{xue2021online2} & 10 & 9.6  & 14.4 & 22.0 \\
    EEND-EDA+FLEX-STB \cite{xue2021online2} & 1  & 13.0 & 16.4 & 23.6 \\
    \textbf{FS-EEND} (prop.)                          & 1  & 10.1 & 14.6 & 21.2 \\
    \midrule
     EEND-EDA+FLEX-STB+VCT \cite{horiguchi2022online} & 1  & 11.1 & 16.0 & 21.7 \\
     \textbf{FS-EEND+VCT} (prop.)                          & 1  & 9.4 & 14.0 & 20.9 \\
    \bottomrule                                                                                                  
\end{tabular}
   \vspace{-1em}
\end{table}

Table \ref{tab:DERs3} shows the results on the CALLHOME data. The BW-EEND-EDA \cite{han2021bw} method is also compared, and the DERs of the best performed configuration are quoted from \cite{han2021bw}. Finetuning the proposed model (by only 250 CALLHOME adaption recordings) with the speaker appearance order does not converge well enough, thus we finetune our model with the PIT loss. The proposed FS-EEND achieves noticeably better results than BW-EEND-EDA and FLEX-STB with 1 s latency, comparable results with FLEX-STB with 10 s latency. Horiguchi et al. promote the performance of FLEX-STB with variable chunk-size training (VCT) in \cite{horiguchi2022online}. We further compare FLEX-STB and the proposed FS-EEND when using VCT for an additional 100 epochs on simulated data and then finetuned on CALLHOME data. Chunk division and optimizer settings remain the same as in \cite{horiguchi2022online}. Note that since we lack the SwitchBoard-2 and 2004 NIST SRE for training data, the results of FLEX-STB are a little worse than the ones reported in \cite{horiguchi2022online} (about 2\%). The results in Table \ref{tab:DERs3} show that the proposed FS-EEND outperforms FLEX-STB when both are promoted with VCT.

\section{Conclusions}
\label{sec:conclu}
This paper proposed a frame-in-frame-out streaming speaker diarization system, which processes the audio stream frame by frame with a causal speaker embedding encoder and an online attractor decoder. With lower inference latency and computational cost, the proposed FS-EEND method achieves better diarization accuracy compared with the block-wise online methods. The current proposed model is not yet suitable for processing very long audio streams, as the self-attention mechanism looks back to all the past frames. This problem will be tackled in our future work.

\vfill\pagebreak
\bibliographystyle{IEEEbib}
\bibliography{Template}

\end{document}